\documentstyle[amssymb,preprint,aps]{revtex}
\begin{document}
\title{$B_{c}${\bf \ meson spectrum and hyperfine splittings in the shifted
large-N-expansion technique }}
\author{Sameer M. Ikhdair$^{\ast }$ and \ Ramazan Sever$^{\dagger }$}
\address{$^{\ast }$Department of Electrical and Electronic Engineering, Near East\\
University, Nicosia, North Cyprus\\
$^{\dagger }$Department of Physics, Middle East Technical University,\\
Ankara,Turkey\\
PACS NUMBER(S): 03.65.Ge, 12.39.Jh, 13.30.Gd}
\date{\today
}
\maketitle

\begin{abstract}
In the framework of potential models for heavy quarkonium, we compute the
mass spectrum of the bottom-charmed $B_{c}$ meson system and spin-dependent
splittings from the Schr\"{o}dinger equation using the shifted-large-N
expansion technique. The masses of the lightest vector $B_{c}^{+}$ and
pseudoscalar $B_{c}$ states as well as the higher states below the threshold
are estimated. Our predicted result for the ground state energy is $%
6253_{-6}^{+15}$ $MeV$ and are generally in exact agreement with earlier
calculations. The parameters of each potential are adjusted to obtain best
agreement with the experimental spin-averaged data (SAD). Our results are
compared with the observed data and with the numerical results obtained by
other numerical methods.
\end{abstract}

\section{INTRODUCTION}

\noindent Recently, theoretical interest has risen in the study of the B$%
_{c} $ meson, the heavy $c\overline{b}$ quarkonium system with open charm
and bottom quarks composed of two nonrelativistic heavy quarks. The spectrum
and properties of the $c\overline{b}$ systems have been calculated various
times in the past in the framework of heavy quarkonium theory [1]. Moreover,
the recent discovery of the lightest vector $B_{c}^{+}$ meson [2] has
inspired new theoretical interest in the subject [3-6]. For the mass of the
lightest vector meson, the CDF Collaboration quotes $M_{B_{c}}=6.40\pm
0.39\pm 0.13$ GeV. This state should be one of a number of states lying
below the threshold for emission of $B$ and $D$ mesons. Such states are very
stable in comparison with their counterparts in $c\overline{c}$ and $b%
\overline{b}$ systems. A particularly interesting quantity should be the
hyperfine splitting that as for c$\overline{c}$ case seems to be sensitive
to relativistic and subleading corrections in $\alpha _{s}$. For the above
reasons it seems worthwhile to give a detailed account of the
Schr\"{o}dinger energies for $c\overline{c}$, $b\overline{b}$ and $c%
\overline{b}$ meson systems below the continuum threshold. Because of the
success of the nonrelativistic potential model and the flavour independence
of the $q_{1}\overline{q}_{2}$ potential, we choose a set of
phenomenological and a QCD-motivated potentials. We insist upon strict
flavor-independence of its parameters. We also use a potential model that
includes running coupling constant effects in both the spherically symmetric
potential and the spin-dependent potentials to give a simultaneous account
of the properties of the $c\overline{c}$, $b\overline{b}$ and $c\overline{b}$
systems. Since one would expect the average values of the momentum transfer
in the various quark-antiquark states to be different, some variation in the
values of the strong coupling constant and the normalization scale in the
spin-dependent potential should be expected.

We have made this study mostly a fully treatment for the potentials used in
the literature. In order to minimize the role of flavor-dependence, we use
the same values for the coupling constant and the renormalization scale for
each of the levels in a given system and require these values be consistent
with a universal QCD scale.

In 1991 Kwong and Rosner $\left[ 7\right] $ predicted the masses of the
lowest vector (triplet) and pseudoscalar (singlet) states of the $B_{c}$
systems using an empirical mass formula and a logarithmic potential. Eichten
and Quigg $\left[ 1\right] $ gave a more comprehensive account of the
energies and decays of the $B_{c}$ system that was based on the
QCD-motivated potential of Buchm\"{u}ller and Tye $\left[ 8\right] .$
Gershtein {\it et al.} $\left[ 9\right] $ also published a detailed account
of the energies and decays of the $B_{c}$ system and used a QCD sum-rule
calculations. Baldicchi and Prosperi [6] have computed the $c\overline{b}$
spectrum based on an effective mass operator with full relativistic
kinematics$.$ They have also fitted the entire light-heavy quarkonium
spectrum. Fulcher $\left[ 4\right] $ extended the treatment of the
spin-dependent potentials to the full radiative one-loop level and thus
included effects of the running coupling constant in these potentials. He
also used the renormalization scheme developed by Gupta and Radford $\left[
10\right] .$

One of the important goals of the present work is to extend the shifted
large-N expansion technique (SLNET) developed for the Schr\"{o}dinger
equation $\left[ 11,12\right] $ to reproduce the $c\overline{b}$
spectroscopy using a class of three static together with Martin and
Logarithmic potentials which was already utilized for producing the
spin-averaged data (SAD) self-conjugate ($q\overline{q}$) mesons and the
(SAD) non-self conjugate ($q_{1}\overline{q}_{2}$) mesons $\left[ 13\right]
. $ We also extend our work using an improved QCD-motivated potential
proposed by Buchm\"{u}ller and Tye $\left[ 8\right] .$

The contents of this article is as follows: in section II, we present the
solution of the Schr\"{o}dinger equation using the SLNET for the the
non-self conjugate $q_{1}\overline{q}_{2}$ mass spectrum. In section III we
present all the potentials used in this work. A brief discussion and
conclusion appear in section IV. \

\section{WAVE EQUATION}

\noindent In this section we shall consider the N-dimensional space
Schr\"{o}dinger equation for any spherically symmetric potential V(r). If $%
\psi ({\bf r})$ denotes the Schr\"{o}dinger's wave function, a separation of
variables $\psi ({\bf r})=Y_{\ell ,m}(\theta ,\phi )u(r)/r^{(N-1)/2}$ gives
the following radial equation (in units $\hbar $=1)[11,12]\bigskip
\begin{equation}
\left\{ -\frac{1}{2\mu }\frac{d^{2}}{dr^{2}}+\frac{[\overline{k}-(1-a)][%
\overline{k}-(3-a)]}{8\mu r^{2}}+V(r)\right\} u(r)=E_{n,\ell }u(r),
\label{1}
\end{equation}
where $\mu =\left( m_{q_{1}}m_{q_{2}}\right) /(m_{q_{1}}+m_{q_{2}})$ is the
reduced mass for the two interacting particles. Here $E_{n,\ell }$ denotes
the Schr\"{o}dinger binding energy, and $\overline{k}=N+2\ell -a,$ with $a$
representing a proper shift to be calculated later on and $\ell $ is the
angular quantum number. We follow the shifted $1/N$ or $1/\overline{k}$
expansion method [12 ] by defining

\begin{equation}
V(r)\;=\frac{\overline{k}^{2}}{Q}\;\left[ V(r_{0})+\frac{V^{\prime
}(r_{0})\;r_{0}\;x}{\bar{k}^{1/2}}+\frac{V^{\prime \prime
}(r_{0})\;r_{0}^{2}\;x^{2}}{2\bar{k}\;}+\;\cdots \right] ,  \label{2}
\end{equation}
and also the energy eigenvalue expansion [12,13]

\begin{equation}
E_{n,\ell }\;=E_{0}+E_{1}/\overline{k}\;+E_{2}/\overline{k}^{2}+E_{3}/%
\overline{k}^{3}+O\left( 1/\overline{k}^{4}\right) ,  \label{3}
\end{equation}
where $x=\overline{k}^{1/2}(r/r_{0}-1)$ with r$_{0}$ is an arbitrary point
where the taylor expansions is being performed about and $Q$ is a scale to
be set equal to $\overline{k}^{2}$ at the end of calculations. Inserting
equations (2) and (3) into equation (1) yields

\[
\left[ -\frac{1}{2\mu }\frac{d^{2}}{dx^{2}}+\frac{1}{2\mu }\left( \frac{\bar{%
k}}{4}-\frac{(2-a)}{2}+\frac{(1-a)(3-a)}{4\;\bar{k}}\right) \right.
\]

\[
\times \left( 1-\frac{2\;x}{\bar{k}^{1/2}}+\frac{3\;x^{2}}{\bar{k}}-\cdots
\right)
\]

\begin{equation}
+\left. \frac{r_{0}^{2}\overline{k}}{Q}\left( V(r_{0})+\frac{V^{\prime
}(r_{0})r_{0}x}{\overline{k}^{1/2}}+\frac{V^{\prime \prime
}(r_{0})r_{0}^{2}\times ^{2}}{2\overline{k}}+\cdots \right) \right] \phi
_{n_{r}}(x)=\xi _{n_{r}}\phi _{n_{r}}(x),  \label{4}
\end{equation}
where the final analytic expression for the $1/\overline{k}$ expansion of
the energy eigenvalues appropriate to the Schr\"{o}dinger particle is [11]
\begin{equation}
\xi _{n_{r}}=\frac{r_{0}^{2}\overline{k}}{Q}\left[ E_{0}+\frac{E_{1}}{\bar{k}%
}+\frac{E_{2}}{\;\bar{k}^{2}}+\frac{E_{3}}{\bar{k}^{3}}+O\left( \frac{1}{%
\overline{k}^{4}}\right) \right] ,
\end{equation}
where $n_{r}$ is the radial quantum number. Here, we formulate the SLNET
(expansion as $1/\bar{k})$ for the nonrelativistic motion of spinless
particle bound in spherically symmetric potential V(r). The resulting
eigenvalue of the N-dimensional Schr\"{o}dinger equation is written as [12]

\begin{eqnarray}
\xi _{n_{r}} &=&\bar{k}\left[ \frac{1}{8\mu }+\frac{r_{0}^{2}V(r_{0})}{Q}%
\right] +\left[ (n_{r}+\frac{1}{2})\;\omega -\frac{(2-a)}{4\mu }\right]
\nonumber \\
&&+\frac{1}{\overline{k}}\left[ \frac{(1-a)(3-a)}{8\mu }+\alpha ^{(1)}\right]
+\frac{\alpha ^{(2)}}{\overline{k}^{2}},  \label{6}
\end{eqnarray}
where $\alpha ^{(1)}$and $\alpha ^{(2)}$are the expressions given by Imbo et
al [11]. Comparing equation (5) with (6) yields

\begin{equation}
E_{0}=V(r_{0})+\frac{Q}{8\mu r_{0}^{2}},  \label{7}
\end{equation}

\begin{equation}
E_{1}=\frac{Q}{r_{0}^{2}}\left[ \left( n_{r}+\frac{1}{2}\right) \omega -%
\frac{(2-a)}{4\mu }\right] ,  \label{8}
\end{equation}

\begin{equation}
E_{2}=\frac{Q}{r_{0}^{2}}\left[ \frac{(1-a)(3-a)}{8\mu }+\alpha ^{(1)}\right]
,  \label{9}
\end{equation}

\begin{equation}
E_{3}=\frac{Q}{r_{0}^{2}}\alpha ^{(2)}.  \label{10}
\end{equation}
Here the quantity $r_{0}$ is chosen so as to minimize the leading term, $%
E_{0}$ [12,13]. That is,

\begin{equation}
\frac{dE_{0}}{dr_{0}}=0\text{ \ \ and \ }\frac{d^{2}E_{0}}{dr_{0}^{2}}>0.
\label{11}
\end{equation}
Therefore, $r_{0}$ satisfies the relation

\begin{equation}
Q=4\mu r_{0}^{3}V^{\prime }(r_{0}),  \label{12}
\end{equation}
and to solve for the shifting parameter $a$, the next contribution to the
energy eigenvalue $E_{1}$ is chosen to vanish [11] so that the second- and
third-order corrections are very small compared with the leading term
contribution. The energy states are calculated by considering the leading
term $E_{0\text{ }}$and the second-order and third-order corrections, it
implies that

\begin{equation}
a=2-2(2n_{r}+1)\mu \omega ,  \label{13}
\end{equation}
with

\begin{equation}
\omega =\frac{1}{4\mu }\left[ 3+\frac{r_{0}V^{\prime
}(r_{0})}{V^{\prime \prime }(r_{0})}\right] ^{1/2}.  \label{14}
\end{equation}
Once $r_{0}$ is being determined, with the choice $\overline{k}=\sqrt{Q}$
which rescales the potential, we derive an analytic expression that
satisfies $r_{0}$ in equations (12)-(14). We have defined the analytic
expression for the $1/\overline{k}$ expansion of the energy eigenvalues (3)
and determined the shifting parameter requiring $E_{1}=0$ $[12,13].$\ \ For
the Coulomb potential, considered as a testing case, the results are found
to be strongly convergent and highly accurate. The calculations of the
energy eigenvalues were carried out up to the second order correction.
Therefore, the Schr\"{o}dinger binding energy to the third order is

\begin{equation}
E_{n,\ell }=E_{0}+\frac{1}{r_{0}^{2}}\left[ \frac{(1-a)(3-a)}{8\mu }+\alpha
^{(1)}+\frac{\alpha ^{(2)}}{\overline{k}}+O\left( \frac{1}{\overline{k}^{2}}%
\right) \right] .  \label{15}
\end{equation}
Once the problem is collapsed to its actual dimension $N=3,$ it simply rests
the task of relating the coefficients of our equation to the one-dimensional
anharmonic oscillator in order to read the energy spectrum. For the $N=3$
physical space, the equation (1) restores its three-dimensional form, one
obtains

\begin{equation}
1+2\ell +(2n_{r}+1)\left[ 3+\frac{r_{0}V^{\prime}
(r_{0})}{V^{\prime \prime }(r_{0})}\right] ^{1/2}=\left[ 8\mu
r_{0}^{3}V^{\prime }(r_{0})\right] ^{1/2}.  \label{16}
\end{equation}
Once $r_{0}$ is being determined through equation (16), the Schr\"{o}dinger
binding energy of the $q_{1}\overline{q}_{2}$ system in equation (15)
becomes relatively simple and straightforward. We finally obtain the total
Schr\"{o}dinger mass binding energy for spinless particles as

\begin{equation}
M(q_{1}\overline{q}_{2})=m_{q_{1}}+m_{q_{2}}+2E_{n,\ell ,}  \label{17}
\end{equation}
where $n=n_{r}+1$ is the principal quantum number. As stated before in
[11-13], for a fixed $n$ the computed energies become more accurate as $\ell
$ increases. This is expected since the expansion parameter $1/N$ or $1/%
\overline{k}$ becomes smaller as $\ell $ becomes larger since the parameter $%
\overline{k}$ is proportional to $n$ and appears in the denominator in
higher-order correction.

\section{SOME POTENTIAL MODELS}

\noindent

\subsection{Static potentials}

The $c\overline{b}$ system that we investigate in the present work is often
considered as nonrelativistic system, and consequently our treatment is
based upon Schr\"{o}dinger equation with a Hamiltonian
\begin{equation}
H_{o}=-\frac{\triangledown ^{2}}{2\mu }+V(r)+V_{SD},  \label{18}
\end{equation}
where $V_{SD}$ is the spin-dependent term having the simple form
\begin{equation}
\text{ \ }V_{SD}\longrightarrow V_{SS}=\frac{32\pi \alpha _{s}}{%
9m_{q_{1}}m_{q_{2}}}\delta ^{3}({\bf r}){\bf s}_{1}.{\bf s}_{2}.  \label{19}
\end{equation}
The spin dependent potential is simply a spin-spin part and this would
enable us to make some preliminary calculations of the energies of the
lowest two S-states of the $c\overline{b}$ system. The potential parameters
in this section are all strictly flavor-independent. The potential
parameters are fitted to the low-lying energy levels of $c\overline{c}$ and $%
b\overline{b}$ systems. The strong coupling constant $\alpha _{s}$ is fitted
to the observed charmonium hyperfine splitting of 117 MeV. The numerical
value of $\alpha _{s}$ is dependent on the potential form and found to be
compatible to the other measurements of reference [1,4,6-7,9]. The hyperfine
observed splitting in charmonium fixes $\alpha _{s}$ for each potential. The
perturbative part of such a quantity was evaluated at the lowest order in $%
\alpha _{s}.$ Baldicchi and Prosperi [6] used the standard running QCD
coupling expression

\begin{equation}
\alpha _{s}({\bf Q})=\frac{4\pi }{\left( 11-\frac{2}{3}n_{f}\right) \ln
\left( \frac{{\bf Q}^{2}}{\Lambda ^{2}}\right) }.  \label{20}
\end{equation}
with $n_{f}=4$ and $\Lambda =0.2~GeV$ cut at a maximum value $\alpha
_{s}(0)=0.35,$ to give the right $J/\psi -\eta _{c}$ splitting and to treat
properly the infrared region. Details on their numerical work are given in
Ref. [6]. Whereas Brambilla and Vairo [3] took in their perturbative
analysis $0.26\leq \alpha _{s}(\mu =2GeV)\leq 0.30.$

The central potentials in (18) include a class of a static potentials of the
general form
\begin{equation}
V(r)=-ar^{-\beta }+br^{\beta }+c\text{, }0<\beta \leq 1,  \label{21}
\end{equation}
previously proposed by Lichtenberg [14] where the parameters $a$ and $b$ are
positive while $c$ may be of either sign. These static quarkonium potentials
are monotone nondecreasing, and concave functions which satisfy the condition

\begin{equation}
V^{\prime }(r)>0\text{ \ and }V^{\prime \prime }(r)\leq 0.  \label{22}
\end{equation}
This comprise the following potentials:\ \ \ \ \ \ \ \ \ \ \ \ \ \ \ \ \ \ \
\ \ \ \ \ \ \ \ \ \ \ \ \ \ \ \ \ \ \ \ \ \ \ \ \ \ \ \ \ \ \ \ \ \ \ \ \ \
\ \ \ \ \ \ \ \ \ \ \ \ \ \ \ \ \ \ \ \ \ \ \ \ \ \ \ \ \ \ \ \ \ \ \ \ \ \
\ \ \ \ \ \ \ \ \ \ \ \ \ \ \ \ \ \ \ \ \ \ \ \ \ \ \ \ \ \ \ \ \ \ \ \ \ \
\ \ \ \ \ \ \ \ \ \ \ \ \ \ \ \ \ \ \ \ \ \ \ \ \ \ \ \ \ \ \ \ \ \ \ \ \ \
\ \ \ \ \ \ \ \ \ \ \ \ \ \ \ \ \ \ \ \ \ \ \ \ \ \ \ \ \ \ \ \ \ \ \ \ \ \
\ \ \ \ \ \ \ \ \ \ \ \ \ \ \ \ \ \ \ \ \ \ \ \ \ \ \ \ \ \ \

\subsubsection{\it Cornell potential}

The QCD-motivated Coulomb-plus-linear potential (Cornell potential) [15]

\begin{equation}
V_{C}(r)=-\frac{a}{r}+br+c,  \label{23}
\end{equation}
with the adjustable parameters as
\begin{equation}
\lbrack a,b,c]=[0.52,0.1756\text{ }GeV^{2},-0.8578\text{ }GeV].  \label{24}
\end{equation}
Such a potential has the Coulomb-like behavior at low distance is supposed
to represent the short range gluon exchange and the term confining the
quarks rises linearly at large distances is supposed to represent the
confinement string tension. The main drawback of this potential is that the $%
c\overline{c}$ and $b\overline{b}$ states lie in an intermediate region of
quark separation where neither limiting forms of (22) should be valid.

\subsubsection{\it Song-Lin potential}

This phenomenological potential was proposed by Song and Lin [16] having the
form

\begin{equation}
V_{SL}(r)=-ar^{-1/2}+br^{1/2}+c,  \label{25}
\end{equation}
with the adjustable parameters

\begin{equation}
\left[ a,b,c\right] =\left[ 0.923\text{ }GeV^{1/2},0.511\text{ }%
GeV^{3/2},c=-0.798\text{\ }GeV\right] .  \label{26}
\end{equation}
The characteristic feature of this potential may be traced in Ref. [16]. \

\subsubsection{\it Turin potential}

Lichtenberg and his co-workers [17] suggested such a potential which is an
intermediate between the Cornell and Song-Lin potentials. This potential
[17] has the form

\begin{equation}
V_{T}(r)=-ar^{-3/4}+br^{3/4}+c\text{,}  \label{27}
\end{equation}
with its adjustable parameters

\begin{equation}
\left[ a,b,c\right] =\left[ 0.620\text{ }GeV^{1/4},0.304\text{ }%
GeV^{7/4},-0.823\text{ }GeV\right] .  \label{28}
\end{equation}

\subsubsection{\it Martin potential}

The phenomenological power-law potential of the form

\begin{equation}
V_{M}(r)=b(r\times 1GeV)^{0.1}+c,  \label{29}
\end{equation}
is labeled as Martin's potential [18] with the parameters values
\begin{equation}
\left[ b,c\right] =\left[ 6.898\text{ }GeV^{1.1},-8.093\text{ }GeV\right] .
\label{30}
\end{equation}
(potential units are also in GeV). \ \ \ \ \ \ \ \ \ \ \ \ \ \ \ \ \ \ \ \ \
\ \ \ \ \ \ \ \ \ \ \ \ \ \ \ \ \ \ \ \ \ \ \ \ \ \ \ \ \ \ \ \ \ \ \ \ \ \
\ \ \ \ \ \

\subsubsection{\it Logarithmic potential}

A Martin's power-law potential reduces [19] into a form

\begin{equation}
V_{L}(r)=b\ln (r\times 1GeV)+c,  \label{31}
\end{equation}
with
\begin{equation}
\left[ b,c\right] =\left[ 0.733\text{ }GeV,-0.6631\text{ }GeV\right] .
\label{32}
\end{equation}
The potential forms in (29), and (31) were used by Eichten and Quigg and all
of these potential forms were also used for $\psi $ and $\Upsilon $ data
probing $0.1$ $fm<r<1$ $fm$\ region.

\subsection{QCD-motivated potentials}

\ \ \ \ \ \ \ \ \ \ \ \ \ \ \ \ \ \ \ \ \ \ \ \ \ \ \ \ \ \ \ \ \ \ \ \ \ \
\ \ \ \ \ \ \ \ \ \ \ \ \ \ \ \ \ \ \ \ \ \ \ \ \

\subsubsection{\it Igi-Ono potential}

The Coulomb-plus-linear potential has been successful in describing the
spectrum of existing quarkonium systems. Moreover, the same is true for
logarithmic potential as well as for power law potentials [8]. These
potentials all describe well the shape of the intermediate region. Neither
the long-distance nor the short-distance nature of the interquark potential
is probed by upsilon and charmonium systems. Other systems depend on the
short- and long- distance parts of the potential as in toponium system. The
potentials for some systems should account properly for the running $\alpha
_{s}$ at the different distances, the $\alpha _{s}$ and the string constant
that describe the long-distance behavior can vary independently.
Furthermore, the $\Lambda _{QCD}$ used to evaluate $\alpha _{s}$ is related
to a specific renormalization scheme so that comparison with other
calculations is possible. This leads to expanding $\alpha _{s}$ to at least
2-loop order and use of nontrivial methods in the interpolation between
long-and short-distance behaviors. The interquark potential at short
distances has been calculated to 2-loop calculations in the modified
minimal-subtraction ($\overline{MS}$) scheme [20]. Together with the 2-loop
expression for $\alpha _{s},$ one has [8,21]

\begin{equation}
V(r)=-\frac{4\alpha _{s}(\mu )}{3r}+\left[ 1+\frac{\alpha _{s}(\mu )}{2\pi }%
(b_{0}ln\mu r+A)\right] ,  \label{33}
\end{equation}
and

\begin{equation}
\alpha _{s}(\mu )=\frac{4\pi }{b_{0}ln(\mu ^{2}/\Lambda _{\overline{MS}}^{2})%
}\left[ 1-\frac{b_{1}}{b_{0}^{2}}\frac{\ln ln(\mu ^{2}/\Lambda _{\overline{MS%
}}^{2})}{ln(\mu ^{2}/\Lambda _{\overline{MS}}^{2})}\right] ,  \label{34}
\end{equation}
with

\begin{equation}
\left[ b_{0,}b_{1,}A\right] =\left[ 11-\frac{2}{3}n_{f},102-\frac{38}{3}%
n_{f},b_{0}\gamma _{E}+\frac{31}{6}-\frac{5}{9}n_{f}\right] .  \label{35}
\end{equation}
Here $n_{f}$ is the number of flavors with mass below $\mu $ and $\gamma
_{E}=0.5772$ is the Euler's number. The renormalization scale $\mu $ is
usually chosen to be $1/r$ to obtain a simple form for $V(r)$. The
singularity in $\alpha _{s}(1/r)$ at $r$=1/$\Lambda _{\overline{MS}}=5$ $%
GeV^{-1}\approx 1$ $fm$ for $\Lambda _{\overline{MS}}=200$ $MeV.$ \ This
singularity can be removed by the substitution

\begin{equation}
\ln \frac{1}{r^{2}\Lambda _{\overline{MS}}^{2}}\longrightarrow f(r)=\ln %
\left[ \frac{1}{r^{2}\Lambda _{\overline{MS}}^{2}}+b\right] .  \label{36}
\end{equation}
The constant $b$ is an adjustable parameter of the potential and will not
affect the perturbative part of the potential. Hence, setting $n_{f}=4$ in
equation (35), the one-gloun exchange part of the interquark potential
simply takes the form

\begin{equation}
V_{OGE}^{(n_{f}=4)}(r)=-\frac{16\pi }{25}\frac{1}{rf(r)}\left[ 1-\frac{462}{%
625}\frac{lnf(r)}{f(r)}+\frac{2\gamma _{E}+\frac{53}{75}}{f(r)}\right] ,
\label{37}
\end{equation}
where the function $f(r)$ can be read off from (36). Furthermore, the long
distance interquark potential grows linearly leading to confinement as
\begin{equation}
V_{L}(r)=ar.  \label{38}
\end{equation}
Igi and Ono [21] proposed a potential whose general form

\begin{equation}
V^{(n_{f}=4)}(r)=V_{OGE}^{(n_{f}=4)}+ar+dre^{-gr},  \label{39}
\end{equation}
so as to interpolate smoothly between the two parts. They added
phenomenologically a term $dre^{-gr}$ to the potential so that to adjust the
intermediate range behavior by which the range of $\Lambda _{\overline{MS}}$
is extended keeping linearly rising confining potential. Hence, the $\Lambda
_{\overline{MS}}$ runs from $100$ to $500~MeV$ keeping a good fit to the $c%
\overline{c}$ and $b\overline{b}$ spectra. Thereby, the potential with $b=20$
is labeled as type I, the one with $b=5$ is labeled as type II, and that one
with $d=0$ and $b=19$ is labeled as type III$.$

\ \ \ \ \ \ \ \ \ \ \ \ \ \ \ \ \ \ \ \ \ \ \ \ \ \ \ \ \ \ \ \ \ \ \ \ \ \
\ \ \ \ \ \ \ \ \ \ \ \ \ \

\subsubsection{\it Improved Chen-Kuang potential}

Chen and Kuang [22] proposed two improved potential models so that the
parameters therein all vary explicitly with $\Lambda _{\overline{MS}}$ so
that these parameters can only be given numerically for several values of $%
\Lambda _{\overline{MS}}.$ Such potentials have the natural QCD
interpretation and explicit $\Lambda _{\overline{MS}}$ dependence both for
giving clear link between QCD and experiment and for convenience in
practical calculation for a given value of $\Lambda _{\overline{MS}}$. It
has the general form
\begin{equation}
V^{(n_{f}=4)}(r)=kr-\frac{16\pi }{25}\frac{1}{rf(r)}\left[ 1-\frac{462}{625}%
\frac{lnf(r)}{f(r)}+\frac{2\gamma _{E}+\frac{53}{75}}{f(r)}\right] ,
\label{40}
\end{equation}
where the string tension is related to Regge slope by $k=\frac{1}{2\pi
\alpha
%TCIMACRO{\UNICODE[m]{0xb4}}%
%BeginExpansion
{\acute{}}%
%EndExpansion
}$. The function $f(r)$ in (40) can be read off from

\begin{equation}
f(r)=ln\left[ \frac{1}{\Lambda _{\overline{MS}}r}+4.62-A(r)\right] ^{2},
\label{41}
\end{equation}
and

\begin{equation}
A(r)=\left[ 1-\frac{1}{4}\frac{\Lambda _{\overline{MS}}}{\Lambda _{\overline{%
MS}}^{I}}\right] \frac{1-\exp \left\{ -\left[ 15\left[ 3\frac{\Lambda _{%
\overline{MS}}^{I}}{\Lambda _{\overline{MS}}}-1\right] \Lambda _{\overline{MS%
}}r\right] ^{2}\right\} }{\Lambda _{\overline{MS}}r},  \label{42}
\end{equation}
with parameters values

\begin{equation}
\left[ k,\alpha
%TCIMACRO{\UNICODE[m]{0xb4}}%
%BeginExpansion
{\acute{}}%
%EndExpansion
,\Lambda _{\overline{MS}}^{I}\right] =\left[ 0.1491\text{ }GeV^{2},1.067%
\text{ }GeV^{-2},180\text{ }MeV\right] .  \label{43}
\end{equation}
The details of this potential can be tracesd in Ref. [22].

\section{DISCUSSION AND CONCLUSION}

We solve the Schr\"{o}dinger equation for various potentials to determine
the position of the $1S$ spin-averaged data (SAD) for $c\overline{c},\ b%
\overline{b},\ $and then for $c\overline{b}$ systems$.\ $\ The hyperfine
splitting of the ground state is given by (19). $\ $The\ hyp$%
%TCIMACRO{\func{erf}}%
%BeginExpansion
\mathop{\rm erf}%
%EndExpansion
$ine\ splitting\ observed\ in\ the\ charmonium family [23]

\begin{equation}
M(J/\psi )-M(\eta _{c})=117MeV,  \label{44}
\end{equation}
fixes the strong coupling constant $\alpha _{s}$ for each potential. For
simplicity and for the sake of comparison we neglect the variation of $%
\alpha _{s}$ with momentum in (20) to have a common spectra for all states
and scale the splitting of $c\overline{b}$ and $b\overline{b}$ from the
charmonium value in (44). The effective constant value, fixed by the
described way, has been applied to the description of not only the $c%
\overline{c}$ system, but also the $b\overline{b},$ and possibly $c\overline{%
b}$ quarkonia. The consideration of the variation of the effective Coulomb
interaction constant becomes especially essential for the $\Upsilon $
particle, for which $\alpha _{s}(\Upsilon )\neq \alpha _{s}(\psi ).$ Thus,
in calculating the splittings of the $c\overline{b}$ spectra, we have to
take into account the $\alpha _{s}(\mu )$ dependence on the reduced mass of
the heavy quarkonium instead of $\alpha _{s}({\bf Q})$ for the reasons
stated in Ref. [3]. That is, the QCD coupling constant $\alpha _{s}$ in
(19)-(20) is defined in the Gupta-Radford (GR) renormalization scheme [10]

\begin{equation}
\alpha _{s}=\frac{6\pi }{\left( 33-2n_{f}\right) \ln \left( \frac{\mu }{%
\Lambda _{GR}}\right) }~,  \label{45}
\end{equation}
in which $\Lambda _{GR\text{ }}$is related to $\Lambda _{\overline{MS}}$ by

\begin{equation}
\Lambda _{GR}=\Lambda _{\overline{MS}}\exp \left[ \frac{49-10n_{f}/3}{%
2\left( 33-2n_{f}\right) }\right] .  \label{46}
\end{equation}
The treatment of our model with momentum dependence form (20) would increase
the accuracy, it probably give very close results and might reproduce the
experimental values equally well within the errors.

Table I reports our prediction for the Schr\"{o}dinger mass spectrum of the
four lowest $c\overline{b}$ $S-$states together with the first three $P-$
and $\ D-$states for various potentials. If we use the $\alpha _{s}$
determined for the $c\overline{c}$ and $b\overline{b}$ systems by other
authors [1,4,6-7,9], we predict their energy masses to within a few $MeV$ to
the calculated SAD [13,17,24]. \ Since the model is spin independent and as
the energies of the singlet states of quarkonium families have not been
measured, a theoretical estimates of these unknown levels introduces
uncertainty into the calculated SAD. Its worthwhile to note that SAD is
defined as the average mass of the $(s=1,\ell =1)$ states in the form $%
SAD(nP_{j})=\frac{1}{9}\left[ 5M(n^{3}P_{2}+3M(n^{3}P_{1}+M(n^{3}P_{0})%
\right] $ and for $(s=1,\ell =0)$ states by $SAD(nS_{j})=\frac{1}{4}\left[
3M(n^{3}S_{1})+M(n^{1}S_{0})\right] $ in which the SAD $S$ level gives the
weight of only $1/4$ to the unknown singlet level and $3/4$ to the known
triplet level.

Instead of showing the calculated masses in $GeV$ as in Ref. [6], it is
useful to report the spectrum for heavy meson masses in $MeV$ obtained by
our formalism in numerical form to four significant figures. Obviously, this
trend provides a measure of the accuracy in reproducing the experimental and
the calculated SAD as one can expect in this match to a nearest few $MeV$.
It has been shown the possibility of producing $c\overline{b}$ mesons in $%
e^{+}e^{-}$ and hadron-hadron colliders [25,26], so that the study of the $c%
\overline{b}$ mesons is not merely academic. It is clear that the
differences between the spectra predicted by different potentials are not
large.

Moreover, the calculated fine and hyperfine splitting values of the vector
and pseudoscalar masses of the $c\overline{b}$ system for the two-lowest $S-$
states\ are also presented in Table I. Predictions for the $c\overline{b}$
ground-state masses depend little on the potential. The $B_{c}$ and $%
B_{c}^{\ast }$ masses and splittings lie within the ranges quoted by Kwong
and Rosner [7] in their survey of techniques for estimating the masses of
the $c\overline{b}$ ground state. Thus our results in Table I are very
similar to those presented in Table I of Eichten and Quigg [1] and also to
that presented by Fulcher [4]. The results obtained with the Song-Lin and
the Martin potential in all cases fall between the extremes defined by the
other potentials. Here, we report the range of the coupling constant we take
in our analysis $0.220\leq \alpha _{s}\leq 0.313$ for all potentials and $%
0.264\leq \alpha _{s}\leq 0.313$ for the class of static potentials in our
study as\ shown in Table I. Overmore, our predictions to the $c\overline{b}$
masses of the lowest S-wave (singlet and triplet) and also their splittings
appear together with those estimated using various methods by many authors
are also shown in Table II. Larger discrepancies among the various methods
occur for the the ground and excited states [6]. As noted in Table II,
Bambilla and Vairo [3] calculated the maximum final result of $\left(
M_{B_{c}^{\ast }}\right) _{pert}=6326_{-9}^{+29}$ $MeV,$ the upper limit
corresponds to the choice of parameters $\Lambda _{\overline{MS}%
}^{n_{f}=3}=350~MeV$ and $\mu =1.2~GeV,$ while the lower limit to $\Lambda _{%
\overline{MS}}^{n_{f}=3}=250~MeV$ and $\mu =2.0~GeV$ as the best
approximation to their perturbative calculation. Its worthwile to note from
Table II, that the SLNET estimations fall in the range demonstrated by other
authors.

The fitted parameters of the potential of type I and II which were proposed
by Igi and Ono are listed in Table III. Our predictions for the $c\overline{b%
}$ mass spectra are reported in Table IV for the type I (with $b=20)$
together with the type II (with $b=5)$. Moreover, the singlet and triplet
masses together with the hyperfine splittings predicted for the two types of
this potential are also reported in Table V. We, hereby, tested acceptable
parameters for $\Lambda _{\overline{MS}}$ from $100$ to $500~MeV$ for the
type I and type II Igi-Ono potentials to produce $\ $the $c\overline{b}$
masses and their splittings. Small discrepancies between our prediction and
SAD experiment [24] can be seen for higher states and such discrepancies are
probably seen for any potential model and it might be related to the
threshold effects or quark-gloun mixings. The fitted set of parameters for
the Igi-Ono potential (type III ) presented by Ref. [21] are also tested in
our method with $b=19$ and $(\Lambda _{\overline{MS}}=300~MeV$ and also $%
390~MeV)$ and then\ $b=16.3$ and $\Lambda _{\overline{MS}}=300~MeV$ which
seems to be more convenient than $\Lambda _{\overline{MS}}=500~MeV$ used by
other authors [8]. Results of this study are also presented in Table VI. It
is clear that the overall study seems likely to be good and the reproduced
masses of states are also reasonable. Once the experimental masses of the $c%
\overline{b}$ spectra becomes available one may be able to sharpen analysis
by adjusting potential parameters and then choosing the mostly convenient
value of $\Lambda _{\overline{MS}}.$ We see that the quark masses $m_{c}$
and $m_{b}$ are sensitive to the variation of $\Lambda _{\overline{MS}},$
their values increase and explained in the following way. The absolute value
of the short-range potential decreases with increasing $\Lambda _{\overline{%
MS}}.$ In order to reproduce experimental masses we need larger quark masses
for larger values of $\Lambda _{\overline{MS}}.$ This situation becomes
completely different if we are allowed to add an additional constant to (39)
and reduce the effect of the exponential term therein for the higher states.
However, we have not also attempted such possibility in this paper for the
sake of comparison with the other authors.

We have found the $q_{1}\overline{q}_{2}$ potentials which reproduce the
experimental masses of the $c\overline{b}$ states for various values of $%
\Lambda _{\overline{MS}}.$ Using this model, we see that the experimental $c%
\overline{b}$ splittings can be reproduced for $\Lambda _{\overline{MS}}\sim
200-400~MeV$ and $b=20$ while $\Lambda _{\overline{MS}}=100$ and $500~MeV$
are clearly ruled out. Taking the preferred value of $\Lambda _{\overline{MS}%
}=400~MeV,$ we can predict the splittings in exact agreement to several $MeV$
with the other formalisms (c.f., Table I of Ref.[6]). We have selected some
preferred values for $\Lambda _{\overline{MS}}$ and $b$ as they provide the
best close fine and hyperfine splittings to the other works as shown in
Table VI.

The predicted $c\overline{b}$ spectra obtained from the Chen-Kuang potential
are also listed in Table VI. We see that for states below the threshold, the
deviations of the predicted spectra from the experimental SAD are within
several $MeV$. We find that $m_{c}$ and $m_{b}$ are insensitive to the
variation of $\Lambda _{\overline{MS}}$ for this potential. This is
consistent with the conventional idea that, for heavy quarks, the
constituent quark mass is close to the current quark mass which is $\Lambda
_{\overline{MS}}$ independent. The $\Lambda _{\overline{MS}}$ dependence of
the Chen-Kuang potential is given in (41)-(42) and the predicted spectra is
obtained for various values of $\Lambda _{\overline{MS}}=100,180$ and $%
375~MeV.$ Numerical calculations show that this potential is insensitive to $%
\Lambda _{\overline{MS}}$ in the range from $100$ to $375~MeV,$ and as $%
\Lambda _{\overline{MS}}$ runs from $375$ to $500~MeV,$the potential becomes
more sensetive. The obtained $n^{1}S_{0}$ and $n^{3}S_{1}$ splittings for
the $c\overline{b}$ system in the Chen-Kuang potential are also listed in
Table VI. For a certain range of $\Lambda _{\overline{MS}}$ the agreement is
good. Moreover, the fine and hyperfine splittings of the S-states in the $c%
\overline{b}$ system predicted by Igi-Ono and Chen-kuang potentials are
listed in Table VII for some proper parameters. They are considerably
smaller than the corresponding values $\Delta _{1S}(c\overline{b})=76$ $MeV,$
and $\Delta _{2S}(c\overline{b})=42$ $MeV$ predicted by the quadratic
formalism. Moreover, Chen-Kuang [22] predicted $\Delta _{1S}(c\overline{b}%
)=49.9$ $MeV,$ and $\Delta _{2S}(c\overline{b})=29.4$ $MeV$ for their
potential with $\Lambda _{\overline{MS}}=200~MeV$ in which the last
splitting is almost constant as $\Lambda _{\overline{MS}}$ increases. Our
predictions for $\Delta _{1S}(c\overline{b})=68$ $MeV,$ and $\Delta _{2S}(c%
\overline{b})=35$ $MeV$ for the Chen-Kuang potential with $\Lambda _{%
\overline{MS}}$ runs from $100$ into $375$ $MeV$. We also find $\Delta
_{1S}(c\overline{b})=67~$ $MeV,$ and $\Delta _{2S}(c\overline{b})=33$ $MeV$
\ for the Igi-Ono potential with $\Lambda _{\overline{MS}}=300$ $MeV$
\bigskip and $b=16.3.$ The present model has the following features: (1) The
present potential predicts smaller $\Delta _{1S}$ and $\Delta _{2S}$ than
the other potentials do for $c\overline{b}\vspace{0.05cm}$ system and the
present $\Delta _{1S}$ and $\Delta _{2S}$ do not depend on $\Lambda _{%
\overline{MS}}$ more sensitively \bigskip (2) The experimental $c\overline{b}
$ spilitting can be repoduced for the preferred range of $\Lambda _{%
\overline{MS}}$ runs from $100$ into $375$ $MeV$. Table VII reports our
results using SLNET compared to other formalisms.

In this paper, we have developed the SLNET in the treatment of the $c%
\overline{b}$ system using group of static and QCD-motivated potentials. For
such potentials the method looks quite attractive as it yields highly
accurate results. The convergence of this method seems to be very fast as
the higher corrections to energy have lower contribution. It is interesting
to note that the scope of this method can be extended to more realistic
potentials.

In this respect, in reproducing the SAD, we used the same fitted parameters
of the other authors for the sake of comparison and also for the sake of
testing the accuracy of our approach. Here, we would expect much better
agreement to experimental data in case of fitting our own parameters
properly. Finally, we comment that the fitted parameters of any potential
are model-dependent in any study.

{\bf Acknowledgement:}One of the authors (S. M. I) gratefully acknowledges
Dr. Suat G\={u}nsel the founder president of the Near East University and
also Prof. Dr. \c{S}enol Bekta\c{s} the vice president for their continuous
support and encourgement.

$\bigskip $ \newpage
\begin{verbatim}

\end{verbatim}

\bigskip $
\begin{array}{l}
\text{TABLE I. The }c\overline{b}\text{ masses and hyperfine splittings (}%
\Delta _{nS}\text{)}^{\ast }\text{ } \\
\text{calculated for some potentials (all in }MeV). \\
\bigskip
\begin{tabular}{lllllll}
\hline\hline
States & [6,24] & Cornell & Song-Lin & Turin & Martin & Logarithmic \\
$\alpha _{s}=$ &  & $0.313$ & $0.264$ & $0.286$ & $0.251$ & $0.220$ \\
$m_{c}~(GeV)=$ &  & $1.840$ & $1.820$ & $1.790$ & $1.800$ & $1.500$ \\
$m_{b}~(GeV)=$ &  & $5.232$ & $5.199$ & $5.171$ & $5.174$ & $4.905$ \\
$M($c$\overline{b})$ &  &  &  &  &  &  \\
$1S$ & $6315$ & $6315$ & $6306$ & $6307$ & $6301$ & $6317$ \\
$1^{3}S_{1}$ & $6334$ & $6335$ & $6325$ & $6326$ & $6319$ & $6334$ \\
$1^{1}S_{0}$ & $6258$ & $6253$ & $6248$ & $6249$ & $6247$ & $6266$ \\
$\Delta _{1S}{}^{\ast }$ & $77$ & $82$ & $76$ & $77$ & $72$ & $68$ \\
$2S$ & $6873$ & $6888$ & $6875$ & $6880$ & $6892$ & $6903$ \\
$2^{3}S_{1}$ & $6883$ & $6897$ & $6884$ & $6889$ & $6902$ & $6911$ \\
$2^{1}S_{0}$ & $6841$ & $6860$ & $6850$ & $6852$ & $6865$ & $6879$ \\
$\Delta _{2S}$ & $42$ & $37$ & $34$ & $36$ & $37$ & $31$ \\
$3S$ & $7246$ & $7271$ & $7209$ & $7246$ & $7236$ & $7225$ \\
$4S$ &  & $7587$ & $7455$ & $7535$ & $7483$ & $7448$ \\
$1P$ & $6772$ & $6743$ & $6733$ & $6731$ & $6730$ & $6754$ \\
$2P$ & $7154$ & $7138$ & $7104$ & $7123$ & $7125$ & $7127$ \\
$3P$ &  & $7464$ & $7371$ & $7428$ & $7398$ & $7375$ \\
$1D$ & $7043$ & $7003$ & $6998$ & $6998$ & $7011$ & $7027$ \\
$2D$ & $7367$ & $7340$ & $7284$ & $7320$ & $7311$ & $7301$ \\
$3D$ &  & $7636$ & $7510$ & $7588$ & $7536$ & $7502$ \\ \hline
\end{tabular}
\\
^{\ast }\Delta _{nS}=M(n^{3}S_{1})-M(n^{1}S_{0}).
\end{array}
$

\bigskip \bigskip

\bigskip $
\begin{array}{l}
\text{TABLE II. The predicted }c\overline{b}\text{ masses of the lowest
S-wave and its splitting } \\
\text{compared with the other authors (all in }MeV). \\
\bigskip
\begin{tabular}{lllllll}
\hline\hline
Work &  & $M_{B_{c}}(1^{1}S_{0})^{\ast }$ &  & $M_{B_{c}^{\ast
}}(1^{3}S_{1}) $ & $\Delta _{1S}$ &  \\
&  &  &  &  &  &  \\
Eichten et al. [1] &  & $6258\pm 20$ &  &  &  &  \\
Colangelo and Fazio [3] &  & $6280$ &  & $6350$ &  &  \\
Baker et al.[27] &  & $6287$ &  & $6372$ &  &  \\
Roncaglia et al. [27] &  &  &  & $6320\pm 10$ &  &  \\
Godfrey et al. [1] &  & 6270 &  & 6340 &  &  \\
Bagan et al. [1,27] &  & 6255$\pm 20$ &  & $6330\pm 20$ &  &  \\
Bambilla et al. [3] &  &  &  & $6326_{-9}^{+29}$ &  &  \\
Baldicchi et al.[6] &  & $6194\sim 6292$ &  & $6284\sim 6357$ & $65\leq
\Delta _{1S}\leq 90$ &  \\
SLNET$^{\dagger }$ &  & $6253_{-6}^{+13}$ &  & $6328_{-9}^{+7}$ & $69\leq
\Delta _{1S}\leq 80$ &  \\
SLNET$^{\ddagger }$ &  & $6258_{-11}^{+8}$ &  & $6333_{-14}^{+2}$ &  &  \\
SLNET &  &  &  & $6310^{\intercal }$ &  &  \\ \hline
\end{tabular}
\\
^{\ast }\text{ The experimental mass of this singlet state is presented in
[2].} \\
^{\dagger }\text{ Averaging over the five values in Table I.} \\
^{\ddagger }\text{ We treat Eichten and Quigg's results [1] in the same
manner.} \\
^{\intercal }\text{ Our best estimation for the center-of-gravity triplet
state.}
\end{array}
$

\bigskip

\bigskip

\bigskip

\bigskip

\bigskip

\bigskip $
\begin{array}{l}
\text{TABLE III. Parameters used for Igi-Ono potential.} \\
\begin{tabular}{llllll}
\hline\hline
$\Lambda _{\overline{MS}}~(GeV)$ & $a~(GeV^{2})$ & $g~(GeV)$ & $d~(GeV^{2})$
& $m_{c}~(GeV)$ & $m_{b}~(GeV)$ \\ \hline
0.1$^{\dagger }$ & $0.1733$ & $.3076$ & $0.4344$ & $1.134$ & $4.563$ \\
0.2 & $0.1587$ & $0.3436$ & $0.2550$ & $1.322$ & $4.731$ \\
0.3 & $0.1443$ & $0.3280$ & $0.0495$ & $1.471$ & $4.868$ \\
0.4 & $0.1387$ & $2.903$ & $0.582$ & $1.515$ & $4.910$ \\
0.5 & $0.1391$ & $2.955$ & $1.476$ & $1.514$ & $4.911$ \\ \hline
0.1$^{\ddagger }$ & $0.1762$ & $0.2753$ & $0.4720$ & $1.120$ & $4.551$ \\
0.2 & $0.1734$ & $0.3479$ & $0.5362$ & $1.267$ & $4.684$ \\
0.3 & $0.1615$ & $0.4482$ & $0.6020$ & $1.416$ & $4.815$ \\
0.4 & $0.1389$ & $0.6219$ & $0.5632$ & $1.604$ & $4.986$ \\
0.5 & $0.1137$ & $1.0029$ & $0.7368$ & $1.748$ & $4.118$ \\ \hline\hline
\end{tabular}
\\
^{\dagger }\text{ Type I potential. } \\
^{\ddagger }\text{ Type II potential. }
\end{array}
$

\bigskip

\bigskip

\bigskip

\bigskip

\bigskip

\bigskip

\bigskip

\bigskip

\bigskip

\bigskip $
\begin{array}{l}
\text{TABLE IV. The c}\overline{b}\text{ mass spectra predicted for various }%
\Lambda _{\overline{MS}}\text{ } \\
\text{using Igi-Ono (type I and II) potential (all in }MeV\text{).} \\
\begin{tabular}{llllllll}
\hline\hline
States & Ref.[6,24] & $\Lambda _{\overline{MS}}=$ & 100 & 200 & $300$ & 400
& 500 \\ \hline
$b=20$ &  &  &  &  &  &  &  \\
$1S$ & $6327$ &  & $6355$ & $6342$ & $6336$ & $6347$ & $6349$ \\
$2S$ & $6906$ &  & $6941$ & $6928$ & $6907$ & $6911$ & $6923$ \\
$3S$ & $7246$ &  & $7290$ & $7266$ & $7270$ & $7272$ & $7274$ \\
$1P$ & $6754$ &  & $6781$ & $6768$ & $6759$ & $6763$ & $6751$ \\
$2P$ & $7154$ &  & $7170$ & $7155$ & $7151$ & $7153$ & $7143$ \\
$1D$ & $7028$ &  & $7055$ & $7041$ & $7030$ & $7031$ & $7019$ \\
$2D$ & $7367$ &  & $7360$ & $7351$ & $7353$ & $7354$ & $7342$ \\
$b=5$ &  &  &  &  &  &  &  \\
$1S$ & $6327$ &  & $6357$ & $6347$ & $6342$ & $6326$ & $6316$ \\
$2S$ & $6906$ &  & $6940$ & $6921$ & $6936^{\ddagger }$ & $6937$ & $6930$ \\
$3S$ & $7246$ &  & $7284$ & $7300$ & $7262$ & $7272^{\dagger }$ & $7238$ \\
$1P$ & 6754 &  & 6782 & 6766 & 6763 & 6749 & 6746 \\
$2P$ & $7154$ &  & $7168$ & $7161$ & $7160$ & $7139$ & $7136$ \\
$1D$ & $7028$ &  & $7055$ & $7038$ & $7038^{\ddagger }$ & $7026$ & $7022$ \\
$2D$ & $7367$ &  & $7361$ & $7346$ & $7340$ & $7335$ & $7347^{\ddagger }$ \\
\hline
\end{tabular}
\\
\begin{array}{l}
^{\dagger }\text{ Carried out to the first order.} \\
^{\ddagger }\text{ Carried out up to the second order correction.}
\end{array}
\end{array}
$

\bigskip

\bigskip

\bigskip

\bigskip

\bigskip

\bigskip

\bigskip

\bigskip

\bigskip

\bigskip

\bigskip

\bigskip

\bigskip

\bigskip

\bigskip

\bigskip

\bigskip

\bigskip

\bigskip

\bigskip

\bigskip

\bigskip

$
\begin{array}{l}
\text{TABLE V. The }c\overline{b}\text{\ mass spectra and }\Delta _{nS}\text{
for various }\Lambda _{\overline{MS}}\text{ } \\
\text{ calculated by using Igi-Ono potential with }\alpha _{s}=0.250\text{
(all in }MeV\text{).} \\
\begin{tabular}{llllllllllll}
\hline\hline
$b=$ &  &  &  & $20$ &  &  &  &  & $5$ &  &  \\ \hline
States & $\Lambda _{\overline{MS}}=$ & 100 & 200 & 300 & 400 & 500 & 100 &
200 & 300 & 400 & 500 \\ \hline
$1S$ & 6327$^{\ast }$ & 6329 & 6316 & 6310 & 6321 & 6323 & 6331 & 6321 & 6316
& 6300 & 6290 \\
$1^{1}S_{0}$ &  & 6276 & 6263 & 6256 & 6268 & 6273 & 6277 & 6271 & 6265 &
6250 & 6239 \\
$1^{3}S_{1}$ &  & 6347 & 6334 & 6328 & 6339 & 6339 & 6349 & 6338 & 6332 &
6317 & 6307 \\ \hline
$\Delta _{1S}$ &  & 71 & 71 & 72 & 72 & 67 & 72 & 67 & 67 & 67 & 68 \\
\hline\hline
$2S$ & 6906$^{\ast }$ & 6915 & 6902 & 6881 & 6885 & 6897 & 6914 & 6895 & 6910
& 6911 & 6904 \\
$2^{1}S_{0}$ &  & 6888 & 6876 & 6856 & 6861 & 6873 & 6888 & 6869 & 6883 &
6883 & 6878 \\
$2^{3}S_{1}$ &  & 6924 & 6911 & 6889 & 6893 & 6905 & 6923 & 6904 & 6920 &
6920 & 6912 \\ \hline
$\Delta _{2S}$ &  & 36 & 35 & 33 & 32 & 32 & 35 & 35 & 37 & 37 & 34 \\
\hline\hline
\end{tabular}
\\
^{\ast }\text{Here we cite Ref. [6].}
\end{array}
$

\bigskip

\bigskip

\bigskip

\bigskip

\bigskip

\bigskip

\bigskip

\bigskip

\bigskip

\bigskip

\bigskip $
\begin{array}{l}
\text{TABLE VI. The }c\overline{b}\text{ mass spectra and }\Delta _{nS}\text{
using} \\
\text{ Igi-Ono (type III) and Chen-Kuang potentials.} \\
\bigskip
\begin{tabular}{lllllll}
\hline\hline
$\Lambda _{\overline{MS}}=$ & $300$ & $300$ & $390$ & $100$ & $180$ & $375$
\\
$b=$ & $16.3$ & $19$ & $19$ &  &  &  \\
$m_{c}=$ & $1.506^{\ast }$ &  &  & $1.478^{\dagger }$ &  &  \\
$m_{b}=$ & $4.897^{\ast }$ &  &  & $4.876^{\dagger }$ &  &  \\
$M($c$\overline{b})$ &  &  &  &  &  &  \\
$1S$ & $6309^{\ddagger }$ & $6337$ & $6298$ & $6323$ & $6323$ & $6323$ \\
$1^{3}S_{1}$ & $6326$ & $6354$ & $6318$ & $6340$ & $6340$ & $6340$ \\
$1^{1}S_{0}$ & $6258$ & $6287$ & $6238$ & $6272$ & $6272$ & $6272$ \\
$\Delta _{1S}{}$ & $67$ & $67$ & $81$ & $68$ & $68$ & $68$ \\
$2S$ & $6880$ & $6898$ & $6878$ & $6879$ & $6879$ & $6879$ \\
$2^{3}S_{1}$ & $6889$ & $6906$ & $6886$ & $6888$ & $6888$ & $6888$ \\
$2^{1}S_{0}$ & $6855$ & $6874$ & $6853$ & $6853$ & $6853$ & $6853$ \\
$\Delta _{2S}$ & $33$ & $33$ & $33$ & $35$ & $35$ & $35$ \\
$3S$ & $7247$ & $7262$ & $7255$ & $7257$ & $7257$ & $7257$ \\
$4S$ & $7553$ & $7569$ & $7564$ & $7570$ & $7570$ & $7570$ \\
$1P$ & $6725$ & $6749$ & $6738$ & $6723$ & $6723$ & $6723$ \\
$2P$ & $7124$ & $7142$ & $7136$ & $7126$ & $7126$ & $7126$ \\
$1D$ & $6997$ & $7018$ & $7014$ & $6992$ & $6992$ & $6992$ \\
$2D$ & $7328$ & $7345$ & $7342$ & $7332$ & $7332$ & $7332$ \\ \hline
\end{tabular}
\\
^{\ast }.\text{Mass (in }GeV)\text{ fitted for Igi-Ono (type III) potential.
} \\
^{\dagger }\text{ Mass (in }GeV)\text{ fitted for Chen-Kuang potential.} \\
^{\ddagger }\text{ Here we cite Ref. [6].}
\end{array}
$

\bigskip

\bigskip

\bigskip\ $\ \ \ \ \ \ \ \ \ \ \ \ \ \ \ \ \ \ \ \ \ \ \ \ \ \ \ \ \ \ \ \ \
\ \ \ \ $

\bigskip

\bigskip

\bigskip

\bigskip

\bigskip

\bigskip

\bigskip

\bigskip $
\begin{array}{l}
\text{TABLE VII. The fine and hyperfine splittings in our work compared with}
\\
\text{that in other works..} \\
\begin{tabular}{llllllll}
\hline\hline
States &  & Quadratic$^{\ast }$ & Linear$^{\ast }$ & Fulcher$^{\ast }$ &
Lattice$^{\ast }$ & C-Q$^{\dagger }$ & I-O$^{\dagger }$ \\ \hline
$b=$ &  &  &  &  &  &  & 16.3 \\
$\Lambda _{\overline{MS}}=$ &  &  &  &  &  & $100-375$ & $300$ \\
Fine splittings &  &  &  &  &  &  &  \\
$M(2S)-M(1S)$ &  & $558^{\ddagger }$ & $533$ & $579$ & $672\pm 120$ & $556$
& $571$ \\
$M(3S)-M(1S)$ &  & $931$ & $899$ &  &  & $934$ & $938$ \\
$M(2P)-M(1P)$ &  & $382$ & $376$ &  &  & $403$ & $399$ \\
$M(2D)-M(1D)$ &  & $324$ & $321$ &  &  & 340 & 331 \\
$b=$ &  &  & 20$^{\dagger }$ & 5$^{\dagger }$ & 19$^{\dagger }$ & 19$%
^{\dagger }$ & 16.3$^{\dagger }$ \\
$\Lambda _{\overline{MS}}=$ &  &  & 400 & 100 & 390 & 300 & 300 \\
$M(2S)-M(1S)$ &  & 558 & 564 & 583 & 580 & 561 & 571 \\
$M(3S)-M(1S)$ &  & 931 & 925 & 927 & 957 & 927 & 938 \\
Hyperfine splittings &  &  &  &  &  &  &  \\
$\Delta _{1S}$ &  & 76 & 62 & 55 & $41\pm 20$ & 68 & 67 \\
$\Delta _{2S}$ &  & 42 & 33 & 32 & $30\pm 8$ & 35 & 33 \\ \hline
\end{tabular}
\\
\begin{array}{l}
^{\ast }\text{ Here we cite Ref.[6].} \\
^{\dagger }\text{ We used SLNET.} \\
^{\ddagger }\text{ Splitting masses (in }MeV)\text{..}
\end{array}
\end{array}
$

\bigskip

\bigskip

\bigskip

\bigskip

\end{document}